\documentclass{desyproc}

\begin{document}
%------------------------------------PHOTON 2009
\newcommand{\zbar}{\bar{z}}
\newcommand{\odd}{\mathbb{O}}
\newcommand{\pom}{\mathbb{P}}
\newcommand{\tmin}{t_{\rm min}}
\newcommand{\smin}{s_{\rm min}}
\newcommand{\gev}{{\rm GeV}}

\title{Looking for the Odderon in photon collisions}

%for single authors the superscripts are optional
\author{{\slshape B. Pire$^1$, F. Schwennsen$^{1,2}$, L. Szymanowski$^3$, and S. Wallon$^2$}\\[1ex]
$^1$CPhT, {\'E}cole Polytechnique, CNRS, 91128 Palaiseau, France \\
$^2$LPT, Universit{\'e} Paris-Sud, CNRS, 91405 Orsay, France \\
$^3$Soltan Institute for Nuclear Studies, Warsaw, Poland}

% if the proceedings are available online (e.g. at Indico)
% please enter the contribution ID or file_name below for the DOI
\contribID{76}
%\contribID{smith\_joe}

% TO THE CONFERENCE EDITORS: 
% please update the following information      
% before sending the template to the authors
\confID{1407}  % if the conference is on Indico uncomment this line
\desyproc{DESY-PROC-2009-03}
\acronym{PHOTON09} % if you want the Acronym in the page footer uncomment this line
\doi  % if there is an online version we will register DOIs

\maketitle

\begin{abstract}
We discuss the production of two pion pairs in photon collisions at high energies. We calculate the according matrix elements in $k_T$-factorization and discuss the possibility to reveal the existence of the perturbative Odderon by charge asymmetries. 
\end{abstract}

\section{Introduction}

At high energies amplitudes of reactions with rapidity gaps in hadronic interactions are dominated by the exchange of a color singlet, $C$-even state  -- called the Pomeron. In the language of perturbative QCD %(pQCD)
 -- at lowest order -- the Pomeron can be described as the exchange of two gluons in the color singlet state. In contrast to the very well settled notion of the Pomeron, the status of its $C$-odd partner -- the Odderon -- is less safe. Although it is needed {\it e.g.} to describe properly  the different behaviors of $pp$ and $\bar p p$ elastic cross sections~\cite{LN}, it still evades confirmation in the perturbative regime, where -- again at lowest order -- it can be described by the exchange of three gluons in the color singlet state.

The main reason lies in its smaller exchange amplitude in comparison to the Pomeron exchange such that in the cross section, obtained after squaring the sum of both amplitudes, the Pomeron amplitude squared dominates. In this contribution we present results of our study~\cite{Pire:2008xe} of a charge asymmetry in the  production of two pion pairs in photon-photon collisions where that Pomeron squared part vanishes. This observable is thus linearly sensitive to the Odderon contribution.

In the present analysis we deal with the hard Pomeron and the hard Odderon 
exchanges, {\it i.e.} both treated within  perturbative QCD. 
This approach can be confronted with a description of the 
Pomeron-Odderon interference based on soft, non-perturbative 
physics and developed in 
Refs.~\cite{Ginzburg:2002zd}. The experimental observation  of the P-O 
interference effects will thus shed a light on the important question 
which of the above mechanisms is more appropriate for the description of data.

\section{Kinematics, amplitudes and GDAs}

\begin{figure}[t]
\centerline{\includegraphics[height=4.9cm]{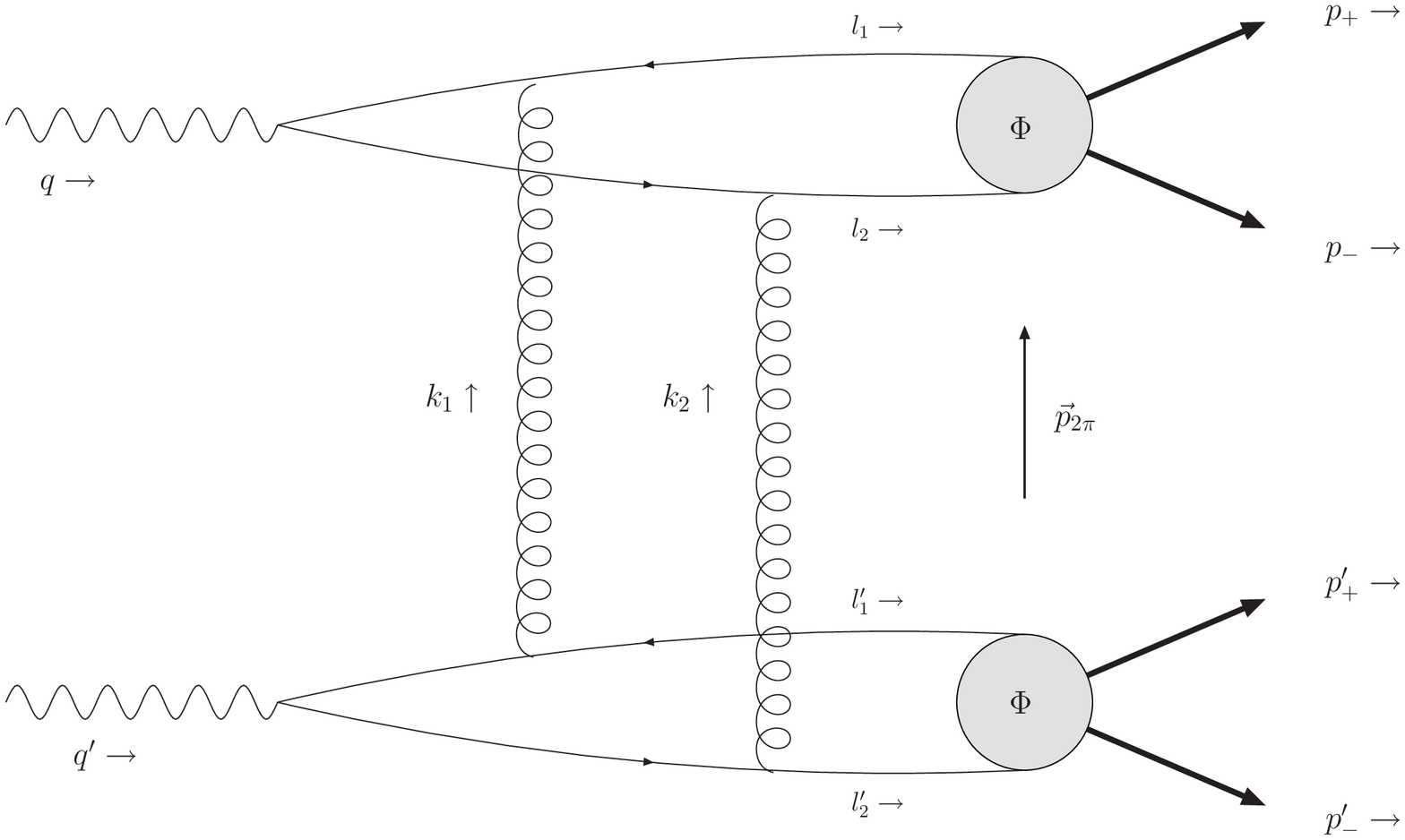}}
\caption{{\protect\small Kinematics of the reaction $\gamma \gamma \to \pi^+ \pi^-\;\; \pi^+ \pi^-$ in a sample Feynman diagram of the two gluon exchange process.}}
\label{fig:1}
\end{figure}

Figure~\ref{fig:1} shows a sample diagram of the process under consideration. We consider large $\gamma\gamma$ energies such that the amplitude can be expressed in terms of two impact factors convoluted over the transverse momenta of the exchanged gluons. The impact factors are universal and consist of a perturbative part -- describing the transition of a photon into a quark-antiquark pair -- and a non-perturbative part, the two pion generalized distribution amplitude (GDA) parametrizing the quark-antiquark hadronization into the 
the pion pair~\cite{Polyakov:1998ze,Diehl:2000uv}.   This comes as a variant of the approach which has been previously proposed in the case of the electroproduction of a pion pair~\cite{Hagler:2002nh,Hagler:2002nf,Warkentin:2007su}, and which is based on the fact that  the $\pi^+ \pi^- $-state does not have any definite charge parity. These GDAs
 which are functions of the longitudinal momentum fraction $z$ of the quark,  of
the angle $\theta$ (in the rest frame of the pion pair) and of the invariant mass $m_{2\pi}$ of the pion pair are the only but nevertheless essential phenomenological inputs. 
In principle, they have to be extracted from experiments but this is a very challenging task and has not been done so far. However, after an expansion in Gegenbauer polynomials $C_n^m(2z-1)$ and in Legendre polynomials $P_l(\beta\cos\theta)$ (where $\beta=\sqrt{1-4m_\pi^2/m_{2\pi}^2}$)~\cite{Polyakov:1998ze}, it is believed that only the first terms give a significant contribution:
\begin{eqnarray*}
  \Phi^{I=1} (z,\theta,m_{2\pi}) &=& 6z\zbar\beta f_1(m_{2\pi}) \cos\theta ,\\
  \Phi^{I=0} (z,\theta,m_{2\pi}) &=& 5z\zbar(z-\zbar)\left[-\frac{3-\beta^2}{2}f_0(m_{2\pi})+\beta^2f_2(m_{2\pi})P_2(\cos\theta)\right],
\end{eqnarray*}
where $f_1(m_{2\pi})$ can be identified with the electromagnetic pion form factor $F_\pi(m_{2\pi})$. 
For the $I=0$ component we use different models. The first model follows Ref.~\cite{Hagler:2002nh} and expresses the functions $f_{0/2}$ in terms of the Breit-Wigner amplitudes of the according resonances.
A second model has been elaborated in Ref.~\cite{Warkentin:2007su} and interprets the functions $f_{0/2}$ as corresponding Omn\`es functions for $S-$ and $D-$waves constructed by dispersion relations from the phase shifts of the elastic pion scattering. 
It has been argued~\cite{Warkentin:2007su,Ananthanarayan:2004xy} that the actual phases of the GDA might be closer to the phases  $\delta_{T,l}$ of the corresponding $T$ matrix elements $\frac{\eta_l e^{2i\delta_l}-1}{2i}$, where $\eta_l$ is the inelasticity factor. The third model for the $I=0$ component of the GDA takes this into account by using the technique of model 2 with these phases $\delta_{T,l}$ of the $T$ matrix elements. Indeed, measurements at HERMES~\cite{Airapetian:2004sy} do not observe a resonance effect at the $f_0$-mass, but concerning the $f_2$ both phases ($\delta_2$ and $\delta_{T,2}$) are compatible with data~\cite{Warkentin:2007su}. Having this in mind, we consider also a fourth model -- a mixed description with the $f_0$ contribution from model 3 and the $f_2$ contribution from model 2.

Recently  the BaBar experiment reported a new measurement of  the reaction $\gamma^\star \gamma \to \pi^0$ 
up to photon virtualities squared of 40~GeV$^2$~\cite{Aubert:2009mc}.
In the latter study, the reaction $\gamma^\star \gamma \to \pi^0 \pi ^0$ was investigated 
in the  $f_0(980)$ and  $f_2(1270)$resonance regions
as a potential background for the study of the $\pi^0$ transition form factor.
This low-$W^2$ kinematical region should soon be analysed in the framework of generalised two-meson distribution
 amplitudes~\cite{VD}. This will settle the question of the adequate model to be used.

\section{Charge asymmetries and rates}

The key to obtain an observable which linearly depends on the Odderon amplitude is the orthogonality of the $C$-even GDA (entering the Odderon process) and the $C$-odd one (entering the Pomeron process) in the space of  Legendre polynomials in $\cos\theta$. Due to an additional multiplication by $\cos\theta$ before the angular integration only the interference term survives. We have to do this for both the pion pairs. Moreover we integrate over the invariant mass of one of the pion pairs to reduce the complexity of our observable. Finally, we define the charge asymmetry in the following way:
\begin{gather*}
 \hat{A}(t,m_{2\pi}^2;m_{\rm min}^2,m_{\rm max}^2) = \frac{\int_{m_{\rm min}^2}^{m_{\rm max}^2} dm_{2\pi}'^2\int\cos\theta\,\cos\theta'\,d\sigma(t,m_{2\pi}^2,m_{2\pi}'^2,\theta,\theta')}{\int_{m_{\rm min}^2}^{m_{\rm max}^2} dm_{2\pi}'^2\int\,d\sigma(t,m_{2\pi}^2,m_{2\pi}'^2,\theta,\theta')}   \\
= \frac{\int_{m_{\rm min}^2}^{m_{\rm max}^2} dm_{2\pi}'^2\int_{-1}^1d\cos\theta\int_{-1}^1d\cos\theta'\;2\cos\theta\,\cos\theta'\,{\rm Re}\left[\mathcal{M}_\pom(\mathcal{M}_\odd+\mathcal{M}_{\gamma})^*\right]}{\int_{m_{\rm min}^2}^{m_{\rm max}^2} dm_{2\pi}'^2\int_{-1}^1d\cos\theta\int_{-1}^1d\cos\theta'\,\left[\left|\mathcal{M}_\pom\right|^2+\left|\mathcal{M}_\odd+\mathcal{M}_{\gamma}\right|^2\right]}
. \label{eq:ahat}
\end{gather*}

\begin{figure}
  \centering
  \includegraphics[width=6cm]{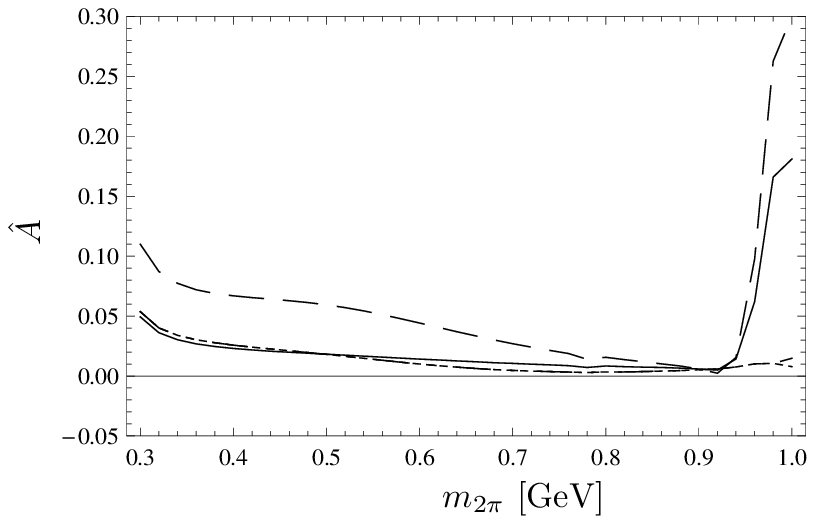}\hspace{10mm}
  \includegraphics[width=6cm]{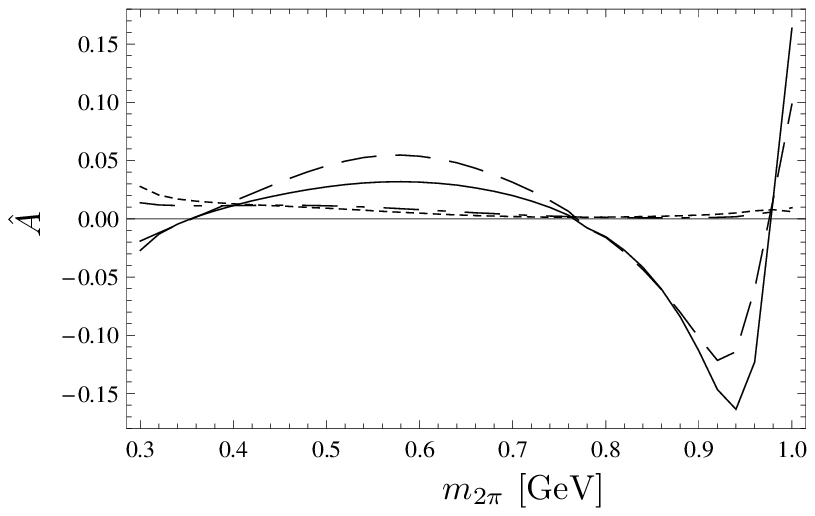}
  \caption{Asymmetry $\hat{A}$ at $t=-1\,\gev^2$ for model 1 (solid), 2 (dashed), 3 (dotted), and 4 (dash-dotted) -- model 3 and 4 are nearly on top of each other. Left column has $m_{\rm min}=.3\,\gev$ and $m_{\rm max}=m_\rho$, while right column has $m_{\rm min}=m_\rho$ and $m_{\rm max}=1\,\gev$. 
}
\label{fig:asymplot1}
\end{figure}

An analytic calculation of the Odderon matrix element would demand the notion of analytic results for two-loop box diagrams, whose off-shellness for all external legs is different. With the techniques currently available such a calculation is not straightforward and we choose to rely on a numerical evaluation by Monte Carlo methods. In particular we make use of a modified  version of {\sc Vegas} as it is provided by the {\sc Cuba} library~\cite{Hahn:2004fe}. The result for $\hat{A}$ is shown in Fig.~\ref{fig:asymplot1} where we took two different choices for the integrated region of the invariant mass of the two pions system. Since our framework is only justified for $m_{2\pi} ^2 < -t$, (in fact strictly speaking, one even needs $m_{2\pi}^2 \ll -t$ ), we keep $m_{2\pi}$ below 1\,GeV. 

\begin{figure}
  \begin{center}
\includegraphics[width=12cm]{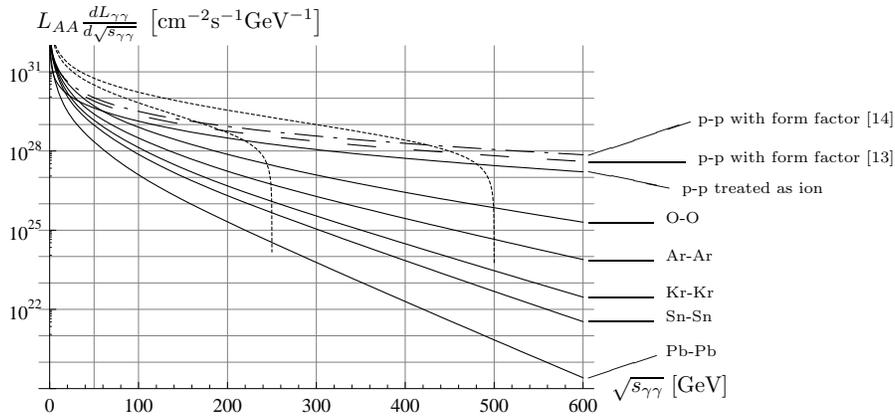} 
  \end{center}
  \caption{Effective $\gamma\gamma$ luminosities for the collision of p-p based on Ref.~\cite{Drees:1988pp} (dash-dotted) and Ref.~\cite{Nystrand:2004vn} (dashed). The results using the parametrization of Ref.~\cite{Cahn:1990jk} for ions are given by solid lines for p-p,
${\rm O}^{8}_{16}$-${\rm O}^{8}_{16}$,
${\rm Ar}^{18}_{40}$-${\rm Ar}^{18}_{40}$,
${\rm Kr}^{36}_{84}$-${\rm Kr}^{36}_{84}$,
${\rm Sn}^{50}_{120}$-${\rm Sn}^{50}_{120}$,
${\rm Pb}^{82}_{208}$-${\rm Pb}^{82}_{208}$ from top to bottom. For ions we used the average luminosities as given in Ref.~\cite{Brandt:2000mu}, for proton we used $L_{pp}=10^{34}\,{\rm cm}^{-2}{\rm s}^{-1}$. For comparison also effective $\gamma\gamma$ luminosities at the ILC are given for $\sqrt{s_{e^+e^-}}=250\,{\rm GeV}$ and $\sqrt{s_{e^+e^-}}=500\,{\rm GeV}$ (both as dotted lines).}
  \label{fig:lumi}
\end{figure}

%\begin{figure}
%  \centering
%  \psfrag{sggingev}{$\begin{matrix}\sqrt{\smin}\\{\rm [GeV]}\end{matrix}$}
%  \psfrag{events}{number of events}
%  \includegraphics[width=10cm]{schwennsen_florian.fig4.eps}
%%\hspace{10mm}\includegraphics[width=6.5cm]{smin2o.eps}
%  \caption{Rate of production of two pion pairs in ultraperipheral collisions for $\tmin=-1\,\gev^2$ in dependence on the lower cut $\smin$ given in `events per month' in case of ions, and `events per six months' in case of protons which in both cases correspond to one year of running of LHC. The solid line displays the result for p-p collision using Ref.~\cite{Drees:1988pp}, the dashed-dotted that for protons treated as heavy ions, the dashed one that for Ar-Ar collisions, and the dotted line that for Pb-Pb collisions. Also the much smaller rates coming from the Odderon exchange are shown (with the same dashing).}
%\label{fig:rates}
%\end{figure}

To answer the question whether it is possible to measure this asymmetry, we adress the two main issues here. First, one has to convolute the $\gamma\gamma$ cross section obtained by our calculations with the effective photon flux at a certain collider. As we discuss in Ref.~\cite{Pire:2008xe} in detail, the most recent review on this topic~\cite{Baltz:2007kq} presents an overview of photon fluxes for different colliding hadrons which is flawed by an  inconsistency in the underlying hadron-hadron luminosities. Therefore, we show a consistent comparison for the design luminosities of different colliding particles in Fig.~\ref{fig:lumi}. 
For the proton case we display the different luminosities based on either the proton form factor~\cite{Drees:1988pp, Nystrand:2004vn} or on the asymptotic formula for large nuclei.

Although photon fluxes are important in p-p collisions at the designed LHC luminosity ($10^{34}\,{\rm cm}^{-2}{\rm s}^{-1}$), data collected under these conditions will suffer from the pile up of events, which will prevent an analysis of the process considered here from being performed. At lower luminosity, rates may be marginally sufficient for values of $-t \approx 1 {\rm GeV}^2$, but designing a trigger strategy to record  interesting events seems very difficult: typical triggers on high $p_t$ mesons demand a minimum $p_T$ of a few GeV, which is incompatible with such low values of $-t $ and the corresponding limit of $ m_{2\pi} < \sqrt{-t}$. Moreover, an important issue is the background in case of hadron colliders. In contrast to electromagnetic processes which have been proposed to be studied in ultraperipheral collisions, pions are produced by pure QCD processes as well.  It is not easy to demonstrate quantitatively that one can separate these two processes by relying on the fact that ultraperipheral processes are strongly peaked at low t, contrarily to the flatter depence of the Pomeron induced ones~\cite{Piotrzkowski:2000rx}.

In nucleus-nucleus collisions, the trigger problem is easily solved by  detecting neutrons from giant dipole resonances in the Zero Degree Calorimeters, but 
the rates are much lower, especially for the heavier ions. The best compromise may be Oxygen-Oxygen collisions, which is by no means the priority of the heavy ion physics community.

In contrast, an electron-positron collider such as the projected ILC would be the ideal environment to study the process under consideration. Photon photon collisions are indeed the dominant processes there and no pile up phenomenon can blur the picture of a scattering event. 
Studies of similar exclusive processes like diffractive double $J/\psi$~\cite{2jpsi} or double $\rho$ production~\cite{2rho} show that high rates are expected.
Maybe an alternative, which we did not study, is a large energy electron ion collider in its ultraperipheral mode.

In conclusion, our proposal to discover the perturbative Odderon through asymmetries in the production of two pion pairs in ultraperipheral collisions at the LHC seems to have a hard time to win over quite severe  experimental constraints. Will the perturbative Odderon continue to escape detection for the next 10 years?

%\section{Conclusion}

%We have shown that in production of pion pairs in $\gamma\gamma$ collisions the charge asymmetry which is linearly dependent on the Odderon amplitude is sizeable and hence offers the possibility to observe the perturbative Odderon in ulraperipheral collisions at the LHC. The concrete values are GDA-model dependent. HERMES measurements of two pion  electroproduction~\cite{Airapetian:2004sy} disfavor models with a strong $f_0$ coupling to  the $\pi^+ \pi^-$ state but to our minds higher statistics data, which  may come from a JLab experiment at 12\,GeV or from the BaBar experiment as already mentioned, are needed  before  a definite conclusion.

\section*{Acknowledgements}

We acknowledge discussions with Mike Albrow, Gerhard Baur, M{\'a}t{\'e} Csan{\'a}d, David d'En\-terria, Bruno Espagnon, Oldrich Kepka, Spencer Klein, Joakim Nystrand, Christophe Royon, and Rainer Schicker.
This work is supported in part by the Polish Grant N202 249235, the French-Polish scientific agreement Polonium, by the grant ANR-06-JCJC-0084 and by the ECO-NET program, contract 12584QK.

%\section{Bibliography}

% ****************************************************************************
% BIBLIOGRAPHY AREA
% ****************************************************************************

\begin{footnotesize}
% IF YOU DO NOT USE BIBTEX, USE THE FOLLOWING SAMPLE SCHEME FOR THE REFERENCES
% ----------------------------------------------------------------------------

% ----------------------------------------------------------------------------

\end{footnotesize}

% ****************************************************************************
% END OF BIBLIOGRAPHY AREA
% ****************************************************************************

\end{document}